\documentclass[11pt]{article}
\usepackage{amsmath}

\usepackage[pctex32]{graphics}

\newcommand{\bee}{\begin{equation}}

\begin{document}

\vspace*{1in}

\begin{center}
{\Large Embedded Solitons in Lagrangian and Semi-Lagrangian Systems}

\vspace{0.5in}

D. J. Kaup$^{1}$ and Boris A. Malomed$^{1,2}$

\vspace{0.2in}

$^{1}$Department of Mathematics\\[0pt]
University of Central Florida\\[0pt]
Orlando, FL 32816, USA\\[0pt]
kaup@ucf.edu\\[0pt]

\vspace{0.2in}

$^{2}$Department of Interdisciplinary Studies\\[0pt]
Faculty of Engineering\\[0pt]
Tel Aviv University\\[0pt]
malomed@eng.tau.ac.il\\[0pt]
\end{center}

{\normalsize \vspace{.1in} }

\vspace{.5in}

\begin{center}
\textbf{Abstract}
\end{center}

We develop the technique of the variational approximation for solitons in
two directions. First, one may have a physical model which does not admit
the usual Lagrangian representation, as some terms can be discarded for
various reasons. For instance, the second-harmonic-generation (SHG) model
considered here, which includes the Kerr nonlinearity, lacks the usual
Lagrangian representation if one ignores the Kerr nonlinearity of the second
harmonic, as compared to that of the fundamental. However, we show that,
with a natural modification, one may still apply the variational
approximation (VA) to those seemingly flawed systems as efficiently as it
applies to their fully Lagrangian counterparts. We call such models, that do
not admit the usual Lagrangian representation, \textit{semi-Lagrangian}
systems. Second, we show that, upon adding an infinitesimal tail that does
not vanish at infinity, to a usual soliton ansatz, one can obtain an
analytical criterion which (within the framework of VA) gives a condition
for finding \textit{embedded solitons}, i.e., isolated truly localized
solutions existing inside the continuous spectrum of the radiation modes.
The criterion takes a form of orthogonality of the radiation mode in the
infinite tail to the soliton core. To test the criterion, we have applied it
to both the semi-Lagrangian truncated version of the SHG model and to the
same model in its full form. In the former case, the criterion (combined
with VA for the soliton proper) yields an \emph{exact} solution for the
embedded soliton. In the latter case, the criterion selects the embedded
soliton with a relative error $\approx 1\%$.

\section{Introduction}

One of the many pioneer contributions by A.C. Newell has been the study of
solitary waves (which we will simply call ``solitons'' here, in line with
currently adopted terminology). These occur not only in integrable models,
but also in many nonintegrable nonlinear-wave systems \cite{N,KN}. It is
toward an improved understanding of solitons in nonintegrable systems, that
we present this work, dedicated to Prof. Newell.

In nonintegrable models, more complex forms of solitons are found as one
considers higher-order systems. An issue of fundamental importance is to
find where, in the space of the soliton parameters, such solutions could
exist. Where this could be is dominantly determined by the model's linear
dispersion relation, $\omega (k)$. Typically, dispersion curves have gaps or
partial gaps, which are values of $\omega $ at which $k$ would take
imaginary or complex values. For example, for the nonlinear Schr\"{o}dinger
(NLS) equation, one has $\omega =Dk^{2}$, where $D$ is a positive dispersion
coefficient, hence $k$ must be imaginary if $\omega <0$. This is precisely
the region where one finds solitons of the NLS equation. For more complex
systems, with several branches of the dispersion relation, one can have
regions of $\omega $ where $k$ can be complex as well as real. Here one
typically encounters ``delocalized solitons'' (see the book \cite{Boyd} and
relevant examples from nonlinear-optical models in Refs. \cite{delocal}),
which are quasi-solitary waves with nonvanishing oscillatory tails. Such
objects, obviously, have an infinite energy, and therefore are unphysical
except, possibly, in finite-size systems. However, it may occur that the
amplitude of the tail vanishes at some special values of $\omega $. Then,
one has a truly localized object at an isolated (discrete) value of $\omega$
at which a real value of $k$ does exist. Since this value of $\omega $ lies
in a continuous part of the spectrum, these objects are called embedded
solitons (ESs). A number of examples of ESs in physically meaningful models
are now known (see examples of such solitons found in the hydrodynamical
models in Refs. \cite{Hydro}, and a short review in Ref. \cite{CMYK01}),
their stability (which turns out to be \textit{semi-stability}) being quite
distinct from the stability of ordinary solitary waves \cite{YMK99}.

Currently, the only method for locating ESs is to search for them
numerically. One does know that they may be found inside a partial gap of
the dispersion relation, but otherwise one has no analytical tool for
locating them (in exceptional cases, exact ES solutions can be found by
guess \cite{YMK99}). The first objective of this paper is to develop an
approximate analytical method for locating ESs. The method is based on the
variational approximation (VA; see a recent review of the application of
this technique to solitons in Ref. \cite{PO}), which incorporates an
amplitude of the infinitesimal tail of the related delocalized soliton as a
key variational parameter. The approach will be tested on two versions of an
ES-generating model introduced in Ref. \cite{YMK99}, which combines the
second-harmonic generation (SHG) through quadratic [$\chi ^{(2)}$]
nonlinearity, and the usual cubic [$\chi ^{(3)}$] nonlinear terms. The
difference between the two versions of the model is that one is a full model
(it has the usual Lagrangian representation), in terms of the $\chi ^{(3)}$
terms, while the one model is a \emph{truncated model}, wherein, upon
assuming that the fundamental-harmonic (FH) field is much stronger than the
second harmonic (SH), some of the SH $\chi ^{(3)}$ terms are omitted. The
truncated model does \textit{not} admit the usual Lagrangian representation.
However, it was this truncated version of the $\chi ^{(2)}:\chi ^{(3)}$
model in which the above-mentioned exact analytic ES solution was found in
Ref. \cite{YMK99}. On the other hand, in the full version of the model, ES
solutions could only be found by means of numerical methods. We will obtain
a natural variational criterion which makes it possible to distinguish ESs
from delocalized solitons in each of these systems. Furthermore, we will
demonstrate that, in the case of the truncated model, this method yields an
exact result, and in the full model, a relative error is $\approx 1\%$ in
the prediction of the location of ES (in comparison with numerical results).

A distinctive feature of the truncated model is that, due to a missing
(omitted) term, it does not have a ``complete'' Lagrangian representation.
Namely, it can be obtained from a Lagrangian, but only if one term in it is
not subjected to the variation with respect to the FH field, when deriving
the system of the FH and SH equations. Accordingly, the truncated model does
not conserve any Hamiltonian. But it is, nevertheless, a conservative
system, as it conserves the norm of the solution (which is usually called
``energy'' in nonlinear optics and is different from the Hamiltonian).

As the truncated model does not admit the full Lagrangian representation, it
is necessary to work out a special version of the VA for it, which is
another objective of this paper. In fact, this can be done in a very simple
way: after inserting the variational \textit{ansatz} into the Lagrangian and
performing the integration in order to obtain the corresponding effective
Lagrangian (the one which is an explicit function of variational parameters,
rather than a functional depending on the field variables), the
above-mentioned special term in the effective Lagrangian should not be
varied with respect to variational parameters that belong to the FH
component of the ansatz. Thus, other models similar to the above-mentioned
truncated one, even though they cannot be represented in the usual
Lagrangian form, can still be handled by means of VA. Since one cannot
freely vary all the fields in the Lagrangian of these systems, we call them 
\textit{semi-Lagrangian} systems.

The rest of the paper is organized as follows. In section 2, we introduce
the full and truncated models and the Lagrangian representations of each. In
section 3, the VA for the semi-Lagrangian case is developed. The general
VA-based analytical condition for identifying ESs (in an approximate form)
is obtained in section 4. Section 5 concludes the paper.

\section{The full and truncated models}

Following Refs. \cite{mixed} and \cite{YMK99}, we first introduce the full 
$\chi ^{(2)}:\chi ^{(3)}$ model: 
\begin{equation}
iu_{z}+(1/2)u_{tt}+u^{\ast }v+\gamma _{1}\left( |u|^{2}+2|v|^{2}\right) u=0,
\label{u}
\end{equation}
\begin{equation}
iv_{z}-(1/2)\delta v_{tt}+qv+(1/2)u^{2}+2\gamma _{2}\left(
|v|^{2}+2|u|^{2}\right) v=0,  \label{v}
\end{equation}
which is written in the usual ``optical'' notation, so that $z$ and $t$ are
the propagation distance and reduced time, $u$ and $v$ are the FH and SH
fields, $-\delta $ is the relative dispersion coefficient at SH, $q$ is the
SHG mismatch, and $\gamma _{1,2}$ are Kerr coefficients. In fact, the ratio
of the self-phase-modulation (SPM) and cross-phase-modulation (XPM)
coefficients is not necessarily $1:2$ simultaneously in both equations, as
it is written in Eqs. (\ref{u}) and (\ref{v}), but this feature of the model
is not a crucially important one.

The form of Eqs. (\ref{u}) and (\ref{v}) implies that the group-velocity
dispersion is anomalous at FH, while at SH it may be both normal, if $\delta
>0$, and anomalous, if $\delta <0$ (both cases are physically possible). As
for the Kerr coefficients $\gamma _{1}$ and $\gamma _{2}$, they always have
one sign. Most typically, they are positive (corresponding to the
self-focusing nonlinearity), but may be negative too, see a detailed
discussion of this point in Ref. \cite{Canberra}.

In many cases, the SH field is much weaker than the FH field - for instance,
if the mismatch is large. Then, assuming that $\left| v\right| ^{2}\ll
\left| u\right| ^{2}$, one may neglect the XPM term in comparison with the
SPM one in Eq. (\ref{u}), and the SPM term in comparison with its XPM
counterpart in Eq. (\ref{v}), which leads to the \textit{truncated model}, 
\begin{equation}
iu_{z}+(1/2)u_{tt}+u^{\ast }v+\gamma _{1}|u|^{2}u=0,  \label{utrunc}
\end{equation}
\begin{equation}
iv_{z}-(1/2)\delta v_{tt}+qv+(1/2)u^{2}+4\gamma _{2}|u|^{2}v=0.
\label{vtrunc}
\end{equation}

In this work, we are interested in stationary fundamental-soliton solutions,
which are looked in the form 
\begin{equation}
u(z,t)=e^{ikz}U(t),\text{ }v(z,t)=e^{2ikz}V(t),  \label{stationary}
\end{equation}
where $k$ is the FH wavenumber, and real even functions $U(t)$ and $V(t)$
with a single maximum at $t=0$ exponentially decay at $t\rightarrow \infty 
$. Note that ordinary (non-embedded) solitons may exist in the regions 
\begin{equation}
0<k<q/2,\,\mathrm{if}\text{ }\delta >0;\,k>\max \{0,q/2\},\mathrm{\,if\,}
\,\delta <0  \label{ordinary}
\end{equation}
(which implies that $q$ must be positive for the existence of ordinary
solitons if $\delta $ is positive). On the other hand, ESs may exist in a
range of $k$ which does not overlap with the continuous spectrum in the FH
equation, (\ref{u}) or (\ref{utrunc}), but falls into the continuous
spectrum of the SH equation, (\ref{v}) or (\ref{vtrunc}). Thus, ES may exist
in the regions [cf. Eqs. (\ref{ordinary})] 
\begin{equation}
k>\max \{0,q/2\},\,\mathrm{if}\text{ }\delta >0;\,0<k<q/2,\mathrm{\,if\,}
\,\delta <0.  \label{ES}
\end{equation}
This implies that, in the case $\delta <0$, the existence of ESs makes it
necessary to have positive $q$.

Substituting the expressions (\ref{stationary}) into Eqs. (\ref{u}) and (\ref
{v}), we arrive at a system of ordinary differential equations, 
\begin{equation}
-kU+(1/2)U^{\prime \prime }+U^{\ast }V+\gamma _{1}\left( U^{2}+2V^{2}\right)
U=0,  \label{U}
\end{equation}
\begin{equation}
-2kV-(1/2)\delta V^{\prime \prime }+qv+(1/2)U^{2}+2\gamma _{2}\left(
V^{2}+2U^{2}\right) V=0,  \label{V}
\end{equation}
where the prime stands for $d/dt$. Accordingly, the system of stationary
equations corresponding to the truncated model is 
\begin{equation}
-kU+(1/2)U^{\prime \prime }+U^{\ast }V+\gamma _{1}U^{3}=0,  \label{Utrunc}
\end{equation}
\begin{equation}
-2kV-(1/2)\delta V^{\prime \prime }+qV+(1/2)U^{2}+4\gamma _{2}U^{2}V=0.
\label{Vtrunc}
\end{equation}

It is obvious that Eqs. (\ref{U}) and (\ref{V}) can be derived from the
Lagrangian 
\begin{equation}
L=\frac{1}{2}\int_{-\infty }^{+\infty }\left[ -kU^{2}-\left( 2k+q\right)
V^{2}-\frac{1}{2}\left( U^{\prime }\right) ^{2}+\frac{\delta }{2}\left(
V^{\prime }\right) ^{2}+U^{2}V+\frac{\gamma _{1}}{2}U^{4}+4\gamma
_{2}U^{2}V^{2}+\frac{\gamma _{2}}{2}V^{4}\right] dt\,.  \label{L}
\end{equation}
In the case of the truncated system of stationary equations, (\ref{Utrunc})
and (\ref{Vtrunc}), one may still obtain this system from the variational
principle, provided we do the following. First, the last term in the
integrand in Eq. (\ref{L}) should be dropped. Second, the next to the last
term should only be subjected to varying in $V$, but not in the $U$ field.

\section{The variational approximation for the semi-Lagrangian system}

In this section, we focus on the application of VA to the semi-Lagrangian
(truncated) system (\ref{Utrunc}), (\ref{Vtrunc}). The most natural
variational ansatz to search for embedded solitons proper (without the tail,
which will be considered in the next section) in this system is based on the
following expressions: 
\begin{equation}
U=A\,\mathrm{sech}\left( \sqrt{2k}x\right) x,\,V=B\,\mathrm{sech}^{2}\left( 
\sqrt{2k}x\right) ,  \label{ansatz}
\end{equation}
where the amplitudes $A$ and $B$ are variational parameters, while the
inverse width $\sqrt{2k}$ has been fixed to match to the linearized form of
Eqs. (\ref{U}) and (\ref{V}) or (\ref{Utrunc}) and (\ref{Vtrunc}) at 
$t\rightarrow \infty $. Substituting the ansatz (\ref{ansatz}) into the
Lagrangian (\ref{L}), dropping the last term, and then performing the
integration, we find the effective Lagrangian of the truncated system: 
\begin{equation}
3\sqrt{2k}L_{\mathrm{eff}}=-4kA^{2}-2\left[ 2\left( 1-\frac{2\delta }{5}
\right) k-q\right] B^{2}+2A^{2}B+\gamma _{1}A^{4}+\frac{35}{5}\gamma
_{2}A^{2}B^{2}.  \label{Leff}
\end{equation}

As it was said above, in order to go from the full system to its truncated
counterpart, one not only has to omit the last term in Eq. (\ref{L}), but
also must take care to not vary the term $4\gamma _{2}U^{2}V^{2}$ with
respect to $U$. In terms of the effective Lagrangian (\ref{Leff}), this
means that, when deriving the variational equations for $A$ and $B$, one
should not vary the last term of the effective Lagrangian with respect to $A$%
. In this case, the variation with respect to $A$ yields a simple equation
which allows one to eliminate $B$, 
\begin{equation}
B=-\gamma _{1}A^{2}+2k.  \label{eliminateB}
\end{equation}
Using this result, the equation produced by varying the Lagrangian with
respect to $B$ can be cast into the final form of a biquadratic equation for 
$A$, 
\begin{equation}
\frac{32}{5}\gamma _{1}\gamma _{2}A^{4}-\left[ 1+\frac{64}{5}k\gamma
_{2}+2\gamma _{1}\left( 2\left( 1-\frac{2\delta }{5}\right) k-q\right)
\right] A^{2}+4k\left[ 2\left( 1-\frac{2\delta }{5}\right) k-q\right] =0.
\label{B}
\end{equation}
Thus, depending on values of the parameters, Eq. (\ref{B}) may give up to
two different physical solutions. The simplest nontrivial case (which still
allows an ES to exist) is with $\gamma _{1}=0$, but $\gamma _{2}\neq 0$.
Then, Eq. (\ref{B}) yields a single solution, 
\begin{equation}
A^{2}=\frac{4k\left[ 2\left( 5-2\delta \right) k-5q\right] }
{5+64k\gamma _{2}},  \label{single}
\end{equation}
which is physical if it gives $A^{2}>0$; the amplitude $B$ can then be
obtained from Eq. (\ref{eliminateB}).

It should be stressed that, as it was demonstrated in Ref. \cite{YMK99}, the
ansatz (\ref{ansatz}) yields exact soliton solutions to Eqs. (\ref{Utrunc})
and (\ref{Vtrunc}) at some uniquely selected value of $k$ [an expression for
it is given below in Eq. (\ref{k})]. Depending on the value of the mismatch
parameter $q$, this exact solution may be either ES or an ordinary soliton.
Comparing Eqs. (\ref{eliminateB}) and (\ref{B}) with that solution, one can
easily verify that expressions (\ref{eliminateB}) and (\ref{B}) are
precisely parts of the exact solution.

Although the exact solution for the ES in the truncated model is available
at the \emph{single} value of $k$ at which it may exist, in the region (\ref
{ordinary}), where a \emph{continuous} family of ordinary solitons is
expected to exist, no general exact solution is known. So, to illustrate the
accuracy and reliability of the modified VA for producing approximate
solutions to the truncated equations (\ref{Utrunc}) and (\ref{Vtrunc}), in
Fig. 1 we display a typical example of a numerically found ordinary soliton
in the region $0<k<q/2$ (in this example, $\delta =1$), together with the
analytical approximation generated by Eqs. (\ref{ansatz}), 
(\ref{eliminateB}), and (\ref{B}). As one looks at the numerical 
solution in Fig. 1, one
observes that the SH component goes slightly negative along the shoulder of
the soliton, and then appears to oscillate as it decays. Note that in the
region where this behavior of the SH component is observed, $U$ is not
small, and, according to Eq. (\ref{vtrunc}), the solution for $V$ in this
region should oscillate indeed. Thus, ordinary solitons in the truncated
system may have fine features which are not found in the ESs.

\section{An analytical criterion to identify embedded solitons}

\subsection{General analysis}

In the case when the wavenumber $k$ falls into the regions (\ref{ES}), ESs
can exist in both the full and truncated models, i.e., (\ref{u}), (\ref{v})
and (\ref{utrunc}), (\ref{vtrunc}) \cite{YMK99}. However, except for using
the exact ES solutions in the truncated model to guess where ESs might exist
in its full counterpart, the only known method for locating ESs in the full
model was to search for them by means of direct numerical computations.

Thus, there is a need for an analytical approach to the quest for ESs. Such
an approach can be based on VA, if one assumes that, at values of $k$ close
to $k_{\mathrm{ES}}$, at which an ES exists, there also exists a family of
delocalized solitons, with small-amplitude oscillating tails in the SH
component that vanish when $k$ become equal to $k_{\mathrm{ES}}$. As it
immediately follows from Eqs. (\ref{V}) or (\ref{Vtrunc}), the free
oscillating tail of a delocalized soliton, which has an infinitesimal
amplitude $b$ and arbitrary phase shift $\psi $, is given by the expression 
\begin{equation}
V_{\mathrm{tail}}=b\cos \left( \sqrt{\left( 2/\delta \right) \left(
2k-q\right) }t+\psi \right) ,  \label{tail}
\end{equation}
that neglects the nonlinear terms. Now, one may add this tail to the ansatz 
(\ref{ansatz}), to have a more general tailed ansatz, 
\begin{equation}
V(t)=V_{\mathrm{sol}}(t)+b\cos \left( \sqrt{\left( 2/\delta \right) \left(
2k-q\right) }t+\psi \right) ,  \label{Vtail}
\end{equation}
where $V_{\mathrm{sol}}(t)$ corresponds to the ansatz for the core of the
delocalized soliton.

As $b$ is an extra variational parameter, one should add the variational
equation 
\begin{equation}
\partial L_{\mathrm{eff}}/\partial b=0  \label{varying}
\end{equation}
to the set of equations obtained by varying the effective Lagrangian with
respect to all the other free parameters (irrespective of the fact if the
system is complete Lagrangian or semi-Lagrangian). As we are interested in
the location of an ES which, by itself, has $b=0$ (no tail), one should set 
$b=0$ \emph{after} completing the differentiation in Eq. (\ref{varying}).
This means that, prior to varying in $b$, one should only keep terms in 
$L_{\mathrm{eff}}$ which are linear in $b$, hence Eq. (\ref{varying}) takes the
general form, 
\begin{equation}
\int_{-\infty }^{+\infty }\left[ \left( \frac{\delta L}{\delta V}\right)
|_{U=U_{\mathrm{sol}}(t),V=V_{\mathrm{sol}}(t)}\right] \cos \left( 
\sqrt{\left( 2/\delta \right) \left( 2k-q\right) }t+\psi \right) dt=0,
\label{general}
\end{equation}
with $\delta /\delta V$ standing for the variational derivative of the
underlying Lagrangian [the one given by Eq. (\ref{L})].

Let us make a couple of observations here. First, the above applies equally
well to the full system as well as to the truncated system, since only
variations in the SH field are involved, while the differences between the
two types of systems involve solely the variation in $U$. Furthermore, since
the solitons sought for are even, the expression $\delta L/\delta V$ with $V$
substituted by $V_{\mathrm{sol}}(t)$ is also even. Hence Eq. (\ref{general})
amounts to an \emph{orthogonality} condition between the infinitesimal tail
and the soliton, 
\begin{equation}
\int_{-\infty }^{+\infty }\left[ \left( \frac{\delta L}{\delta V}\right)
|_{U=U_{\mathrm{sol}}(t),V=V_{\mathrm{sol}}(t)}\right] \cos \left( \sqrt{
\left( 2/\delta \right) \left( 2k-q\right) }t\right) dt=0.  \label{cos}
\end{equation}
This consideration also shows that the phase parameter $\psi $ in the
expression (\ref{tail}) is not important in the limit of $b=0$.

We notice that the above derivation circumvents the formal problem of the
divergence of the integral expression (\ref{L}) for the Lagrangian, when
there is a tail which does not vanish as $\left| t\right| \rightarrow \infty 
$. The divergence did not appear since the tail was designed as a solution
to the linearized version of Eqs. (\ref{V}) or (\ref{Vtrunc}). One can
readily verify that the divergence can also be eliminated in the fully
nonlinear case, with respect to the tail (provided one suitably adjusts the
frequency of the tail according to the nonlinearity).

Thus, Eq. (\ref{cos}) is a general criterion that can be used to locate ES
solutions within the framework of VA. In Eq. (\ref{cos}), we recognize that
the variational derivative $\left( \delta L/\delta V\right) |_{U=U_{\mathrm{
sol}},V=V_{\mathrm{sol}}(t)}$ is just the left-hand side (l.h.s.) of the
stationary equation for SH, with $U$ and $V$ taken as per the chosen ansatz
for the core of the delocalized soliton. Another point is that upon applying
integration by parts to the second-derivative term in Eq. (\ref{Vtrunc}),
one sees that all contributions from the linear terms cancel in Eq. (\ref
{cos}), leaving only the nonlinear terms in Eq. (\ref{Vtrunc}) to determine
this condition.

\subsection{Embedded soliton in the truncated system}

To test the efficiency of the criterion (\ref{cos}), we first apply it to
the truncated system. Then, the variational derivative in Eq. (\ref{cos}) is
replaced by the nonlinear part of l.h.s. of Eq. (\ref{Vtrunc}), 
\begin{equation}
\int_{-\infty }^{+\infty }\left[ \frac{1}{2}U^{2}(t)+4\gamma _{2}U^{2}(t)V(t)
\right] \cos \left( \sqrt{\left( 2/\delta \right) \left( 2k-q\right) }
t\right) dt=0.  \label{k1}
\end{equation}
Using the ansatz (\ref{ansatz}), it is easy to explicitly perform the
integration in Eq. (\ref{k1}), which finally yields a simple result (note
that the FH amplitude drops out), 
\begin{equation}
4\gamma _{2}B=-\frac{3\delta \cdot k}{2k\left( 1+2\delta \right) -q}\,.
\label{k2}
\end{equation}
Now, combining the above results (\ref{B}) and (\ref{eliminateB}), which
were obtained by means of the VA, with the relation (\ref{k2}) that locates
where the ES must be (and is also based on VA), one can easily verify that
this set of three relations is \emph{precisely tantamount} to the exact
analytical solution for the ES which was found by guess in Ref. 
\cite{YMK99}. In particular, an eventual expression for the wavenumber of ES is 
\begin{equation}
k_{\mathrm{ES}}=\frac{1}{2}\left( 1+2\delta \right) ^{-1}\left[ q-\frac{3}{2}
\delta \left( 4\gamma _{2}+3\delta \gamma _{1}\right) ^{-1}\right].
\label{k}
\end{equation}

The fact that, in the case of the semi-Lagrangian truncated system, VA
reproduces the \emph{exact} ES is remarkable, although the reason for this
occurrence is not fully understood. We also note that this exact soliton is
not always an embedded one, as the wavenumber (\ref{k}) is not necessarily
restricted to the region (\ref{ES}) in which ES may exist: the wavenumber
may instead fall into the region (\ref{ordinary}), in which ordinary
solitons are to be found. A condition (inequality) showing whether the exact
soliton is embedded or ordinary has already been given in Ref. \cite{YMK99}.

\subsection{Embedded soliton in the full system}

The next step is to apply the general ES-selecting criterion (\ref{cos}) to
the full system which is based on Eqs. (\ref{U}) and (\ref{V}). To this end,
we assume that the soliton proper may again be approximated by the ansatz 
(\ref{ansatz}). Then, substituting the nonlinear part of l.h.s. of Eq. (\ref
{V}) into Eq. (\ref{cos}), we arrive, instead of Eq. (\ref{k2}), at a more
complicated relation, 
\begin{equation}
4\gamma _{2}B\left[ 1+\frac{B^{2}}{A^{2}}\frac{2\left( 8\delta +1\right) k_
{\mathrm{ES}}-q}{40\delta \cdot k_{\mathrm{ES}}}\right] =-\frac{3\delta \cdot
k_{\mathrm{ES}}}{2k_{\mathrm{ES}}\left( 1+2\delta \right) -q}\,.  \label{k3}
\end{equation}
We test the validity of this relation in the following way: take a
particular example of the ES in the full system that was found in a
numerical form in Ref. \cite{YMK99}, for which $\delta =1$, $q=1$, and 
$\gamma _{1}=\gamma _{2}=-0.05$. Borrowing values of the amplitudes $A$ and $B
$ directly from the numerical data, we find that $A=3.794$ and $B=2.735$.
Substituting these values into Eq. (\ref{k3}) yields $k_{\mathrm{ES}}=0.688$%
, while the numerical value found in Ref. \cite{YMK99} was $k_{\mathrm{ES}%
}=0.696$. Thus,the relative error of the criterion (\ref{cos}) for this case
is $1.1\%$.

\section{Conclusion}

In this work, we have put forward two modifications to the technique of the
variational approximation for solitons. First, it may happen that a physical
model does not admit the full Lagrangian representation, as some terms may
be missing due to various reasons. In the case of the $\chi ^{(2)}:\chi
^{(3)}$ model considered in this work, this means that one term in the
Lagrangian should not be varied when deriving the equation for the
fundamental wave. We demonstrate that the VA can be applied to such \textit{
semi-Lagrangian} systems as efficiently as to their full Lagrangian
counterparts. Second, we have shown that, by the addition of an
infinitesimal tail, which does not vanish at infinity, to the usual soliton
ansatz, and demanding, after performing the variation, that the amplitude of
the tail be zero, we obtain an approximate analytical criterion for locating
embedded solitons inside a family of delocalized ones, i.e., isolated truly
localized solutions existing inside the continuous spectrum of radiation
modes. The criterion takes the form of orthogonality between the radiation
mode contained in the infinitesimal tail and the core of the delocalized
soliton. To test the criterion, we have applied it to both the
semi-Lagrangian truncated version of the $\chi ^{(2)}:\chi ^{(3)}$ model and
to the same model in its full form. In the former model, the criterion,
combined with the VA for the soliton proper, yields a result which
completely coincides with the previously found exact solution for the
embedded soliton. In the latter model, the criterion predicts the wavenumber
corresponding to the embedded soliton with a relative error $\approx 1\%$.

\section*{Acknowledgement}

B.A.M. appreciates hospitality of the Department of Mathematics at the
University of Central Florida. This research has been supported in part by
NSF grant \# DMS0129714.

\section*{Figure Captions}

Fig. 1. A comparison between the numerical shape (solid curves) of an
ordinary (non-embedded) soliton of the semi-Lagrangian (truncated) system of
equations (\ref{Utrunc}) and (\ref{Vtrunc}), as found by the shooting
method, and the result (dashed curves) obtained from the modified
variational approximation based on Eqs. (\ref{ansatz}), (\ref{eliminateB}),
and (\ref{B}). The parameters are $q=1$, $\delta =1$, $\gamma _{1}=0$, and 
$\gamma _{2}=-1/4$, and both the numerical and analytical solutions are taken
for $k=1/4$.

\vspace{0.3in}

\hfil\scalebox{0.9}{\includegraphics{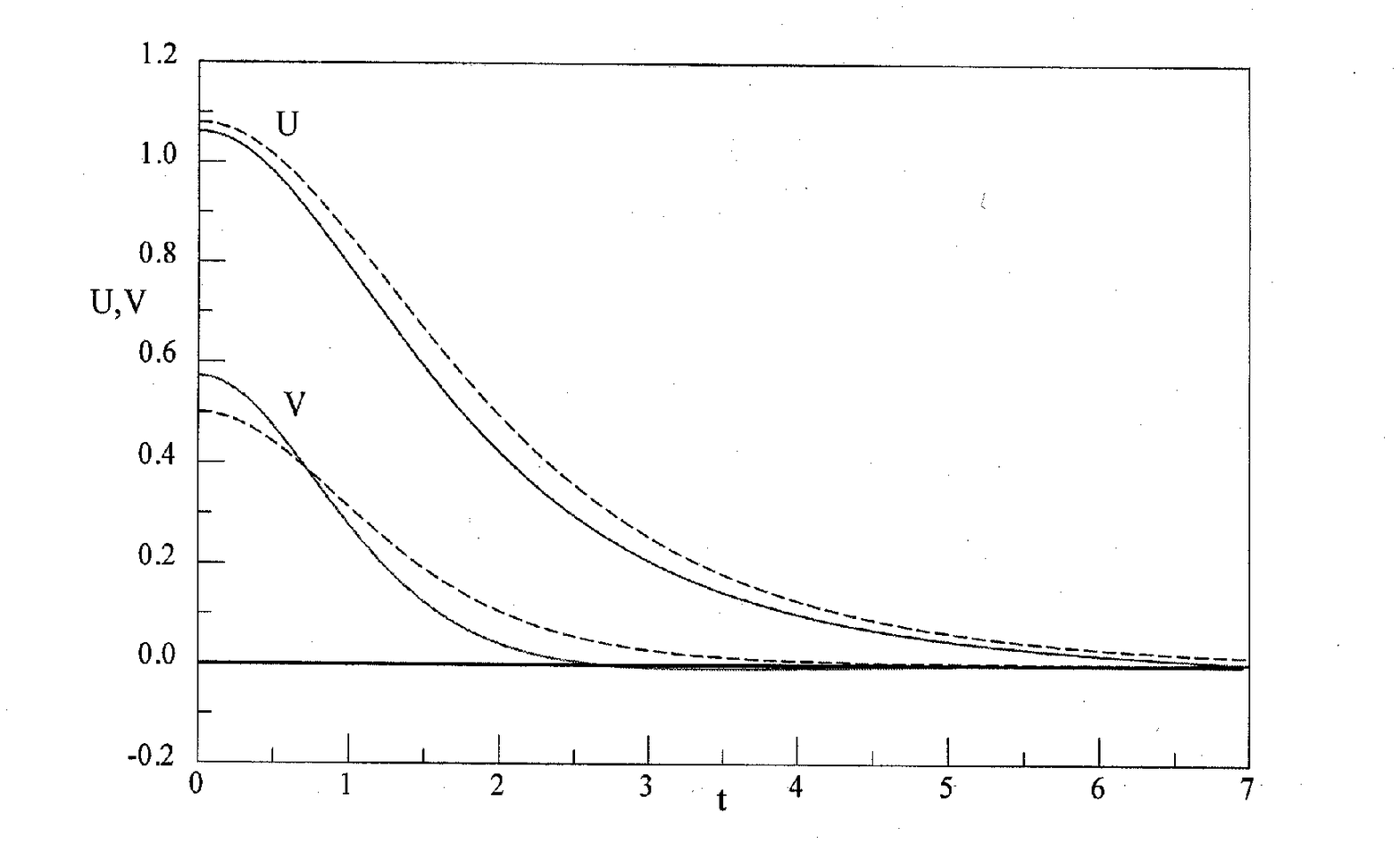}}\hfil

\end{document}